\documentclass[aps,pra, reprint, amsmath, amssymb, groupedaddress]{revtex4-1}

\usepackage{graphicx, subfigure, dcolumn}
\usepackage{bm}

\begin{document}

\title{Laser cooling of high temperature oscillator by multi-level system}

\author{Hoi-Kwan Lau}\email[Email address: ]{hklau.physics@gmail.com}
\affiliation{Institute of Theoretical Physics, Ulm University, Albert-Einstein-Allee 11, 89069 Ulm, Germany}
\author{Martin B. Plenio}
\affiliation{Institute of Theoretical Physics, Ulm University, Albert-Einstein-Allee 11, 89069 Ulm, Germany}

\date{\today}

\begin{abstract}
We study the laser cooling of a mechanical oscillator through the coupling with a dissipative three-level system.  
Under a background temperature beyond the Lamb-Dicke regime, we extend the standard cooling analysis by separately studying the classical motion and the quantum dynamics of the oscillator.  
In Ladder-system cooling, the cooling rate degrades by orders of magnitude at large classical motion.  This phenomenon causes a critical transition of the final temperature at a hot background.  
In stark contrast, electromagnetic-induced-transparency (EIT) cooling with a $\Lambda$-system produces significant negative cooling rate at high motional excitation.  At steady state, the oscillator could exhibit both cooling and lasing behaviours.  We argue that a successful EIT cooling requires either a poor quality oscillator to suppress the lasing effect, or terminating the cooling process at a transient stage.
\end{abstract}


\maketitle

\section{Introduction\label{sec:Intro}}

A macroscopic mechanical oscillator with quantum-level motion is of great interest due to its wide range of applications in establishing quantum information channels and storage \cite{Cleland:2004jm,Rabl:2010kk,Habraken:2012uc}, testing the foundation of quantum mechanics \cite{Ghirardi:1986ep,Ghirardi:1990ej,Marshall:2003kj}, measuring forces as weak as the Casmir force \cite{Bordag:2001be}, and many others \cite{Schwab:2005fy}.  By using state-of-the-art refrigeration technology, a GHz range oscillator can be cooled to the single phonon level under a background temperature $\sim100$ mK \cite{OConnell:2010br}.  In some applications, however, a cooled oscillator with lower frequency is more desirable due to its larger zero point motion and the slower control requirement.  Even at cryogenic temperature, the thermal occupation of an oscillator with frequency $\lesssim$100MHz is much larger than unity.  Motional excitation has to be removed by applying additional cooling processes, such as feedback cooling \cite{Hopkins:2003hy,Poot:2012fh}, sideband cooling \cite{WilsonRae:2007jp,Marquardt:2007dn,Teufel:2011jg, Chan:2011dy}, ultrafast pulsed laser cooling \cite{Machnes:2010bg, Machnes:2012bg}, or laser cooling with a dissipative finite-level (qudit) system \cite{Martin:2004kv, WilsonRae:2004il, Xia:2009kl, Rabl:2010cm, Kepesidis:2013fv}.

The latter method is particularly favourable for quantum information processors because the dissipative qudit can be implemented by the inherent quantum memory initialization mechanism.  
If the initial temperature of the oscillator is sufficiently low, i.e., within the Lamb-Dicke (LD) regime, the performance of laser cooling can be analysed by using the techniques developed in atomic systems \cite{WilsonRae:2004il}.  In this regime, a realistic mechanical oscillator could be cooled to near the motional ground state through the coupling with a quantum dot \cite{WilsonRae:2004il}, a superconducting circuit \cite{Jaehne:2008gw,Xia:2009kl}, or a diamond NV centre \cite{Rabl:2009fz}.  

In practice, however, the initial equilibrium temperature of a mechanical oscillator is often far beyond the LD regime.  The high temperature performance of a two-level-system (TLS) laser cooling have been studied by computing the transition matrix of the highly excited states \cite{Martin:2004kv}, or considering an expansion of the derivatives of the quasi-probability distribution \cite{Rabl:2010cm, Kepesidis:2013fv}.  The cooling rate is generally found to drop significantly at high motional excitation.  Nevertheless, the final excitation predicted by the LD regime analysis is valid even beyond the LD regime.


In some architecture, the oscillator is coupled with a system, such as a diamond NV centre, that intrinsically involves more than two internal levels.  The performance of laser cooling could be improved by utilising the additional states.  In this article, we specifically study the the laser cooling of a high temperature oscillator with a three-level system.
The objective of our work is three-fold: first, we present a strategy to study the evolution of a highly excited oscillator by separating its classical motion from its quantum dynamics.  Our method is an extension of the LD regime analysis, and assumes only the regularity of the quasi-probability distribution but not the magnitude of its derivatives.  Second, TLS cooling is efficient only when the system parameters are appropriate.  In the low temperature regime, desired effective parameters can be engineered by appropriately driving a multi-level system, such as a Ladder-system \cite{Marzoli:1994tz}.  Our work shows that in spite of the discrepancy of cooling rate in the high temperature regime, the final excitation produced by a Ladder-system is close to that of an effective TLS when cooling is efficient.  Third, quantum coherence between multiple metastable states can introduce cooling effects that behave differently from TLS cooling.  Particularly, 
a lower final excitation can be achieved by using electromagnetic-induced-transparency (EIT) cooling with a $\Lambda$-system \cite{Morigi:2000wv,Roos:2000ch,Morigi:2003cv,Lin:2013be, Lechner:2016EIT}.
In the high temperature regime, we find that EIT cooling generally exhibits both cooling and phonon-lasing effects.  Modifications of the cooling process would be required for EIT cooling to be efficient.

Our article is outlined as follows.  In Sec. \ref{sec:LD_cooling}, we briefly review the theoretical analysis of laser cooling in the LD regime for future comparison.  In Sec. \ref{sec:classical_displacement}, we derive an effective Fokker-Planck equation to describe the dynamics of the Glauber P distribution during laser cooling under a high temperature background.  In Secs. \ref{sec:Ladder} and \ref{sec:Lambda}, we apply our tools to analysis the cooling performance of a Ladder- and a $\Lambda$-system respectively.  We conclude the article in Sec. \ref{sec:conclusion}.  In this article, we attribute the subscript $q$ ($a$) to the quantities belonging to the qudit (oscillator mode), and we have chosen units such that $\hbar=1$.

\section{Cooling in Lamb-Dicke regime \label{sec:LD_cooling}}

In this section, we briefly review the analysis of laser cooling in the LD regime.  More details can be found in Refs. \cite{Cirac:1992tp, Jaehne:2008gw, Rabl:2010cm}.  We consider a mechanical oscillator is coupled with a system with multiple internal levels.  By selecting appropriate system parameters, we can focus on the evolution of only one oscillation mode and a $d$-level system (qudit).  The dynamics of the total mode-qudit state, $\rho$, is governed by the master equation,
\begin{equation}\label{eq:master_LD}
\dot{\rho} = \mathcal{L}_q\rho +  \mathcal{L}_\textrm{int}\rho + \mathcal{L}^0_a \rho+ \mathcal{L}^\mathcal{D}_a\rho~.
\end{equation}
The first part accounts for the unperturbed evolution of the qudit, $\mathcal{L}_q\rho \equiv -i[H_q,\rho] + \mathcal{L}^\mathcal{D}_q\rho $, where $H_q$ is the Hamiltonian of the qudit.  $\mathcal{L}^0_a \rho \equiv - i [\nu \hat{a}^\dag \hat{a},\rho ] $ accounts for the unperturbed evolution of the oscillation mode with the annihilation operator $\hat{a}$ and the oscillation frequency $\nu$.  $ \mathcal{L}_\textrm{int}\rho= -i[H_\textrm{int},\rho]$ accounts for the interaction between the mode and the qudit.  The lowest order mode-qudit interaction is usually linear to the mode operators, i.e.,
\begin{equation}\label{eq:H_int}
H_\textrm{int}=\lambda V (\hat{a}+\hat{a}^\dag)~, 
\end{equation}
where $V$ is an operator acting on only the qudit levels.  The effective Lamb-Dicke parameter is defined as the ratio of the interaction strength to the mode frequency, i.e., $\eta \equiv \lambda/ \nu$ \cite{WilsonRae:2004il}.  $\mathcal{L}^\mathcal{D}_q$ and $\mathcal{L}^\mathcal{D}_a$ account for the open-system dynamics of the qudit and oscillation mode respectively.  In particular,
\begin{equation}
\mathcal{L}^\mathcal{D}_a \rho = \gamma (N_\textrm{th}+1) \mathcal{D}[\hat{a}]\rho +  \gamma N_\textrm{th} \mathcal{D}[\hat{a}^\dag]\rho~,
\end{equation}
where $\mathcal{D}[\hat{o}]\rho\equiv \hat{o}\rho \hat{o}^\dag - (\hat{o}^\dag \hat{o}\rho+\rho\hat{o}^\dag \hat{o})/2$ for some operator $\hat{o}$; the bare damping rate $\gamma$ is defined from the Q factor, i.e., $\gamma\equiv\nu/Q$, where $Q$ is the number of cycle to lose $1-1/e$ of motional energy to a zero temperature background.  $N_\textrm{th}\equiv k_B T/\hbar \nu$ is the mean phonon number of the mode if the oscillator is in equilibrium with a background with temperature $T$.

The LD regime considers the scenario of a weak interaction strength, i.e., $\eta \ll 1$, and the equilibrium phonon number satisfying the criterion
\begin{equation} \label{eq:LD_criterion}
\eta \sqrt{N_\textrm{th}} \ll1~.
\end{equation}
In this regime, the mode-qudit interaction described by $H_\textrm{int}$ is always a perturbative effect.  In the master equation Eq.~(\ref{eq:master_LD}), the unperturbed dynamics is governed by $\mathcal{L}_q + \mathcal{L}^0_a$; the interaction $\mathcal{L}_\textrm{int}$ is a first order effect; and the mode dissipation $\mathcal{L}^\mathcal{D}_a$ is assumed to be in the second order.

At the leading order, standard laser cooling analysis considers 
 the system could be represented by a separable state $\rho_a \otimes \rho_{ss}$.  The qudit steady state $\rho_{ss}$ can be obtained self-consistently by solving
\begin{equation}\label{eq:ss_q}
\dot{\rho}_\textrm{ss}=\mathcal{L}_q \rho_\textrm{ss} -i [\lambda V (\alpha_\textrm{ss}+\alpha^*_\textrm{ss}), \rho_\textrm{ss} ]=0~.
\end{equation}
The steady state displacement, $\alpha_\textrm{ss}$, induced by the qudit steady state follows the relation
\begin{equation}\label{eq:ss_a}
(i \nu + \gamma/2) \alpha_\textrm{ss} + i \lambda \langle V\rangle_\textrm{ss}=0~,
\end{equation}
where $\langle V\rangle_\textrm{ss}\equiv \textrm{Tr}\{V\rho_\textrm{ss} \}$.
The steady state contribution can be subtracted off by considering the displaced mode $\tilde{\rho}_a \equiv D^\dag(\alpha_\textrm{ss})\rho_a D(\alpha_\textrm{ss})$, where $\rho_a\equiv \textrm{Tr}_q\{\rho\}$.

By adiabatically eliminating the qudit contribution, the quantum dynamics of the mode state follows
\begin{eqnarray}\label{eq:ssmode_eq}
\dot{\tilde{\rho}}_a &=& \mathcal{L}^0_a \tilde{\rho}_a + \mathcal{L}^\mathcal{D}_a \tilde{\rho}_a  \\ 
&&+ \textrm{Tr}_q \Big\{ \int_0^t \tilde{\mathcal{L}}_\textrm{int} e^{(\mathcal{L}_q +\mathcal{L}^0_a ) \tau}  \tilde{\mathcal{L}}_\textrm{int} e^{-\mathcal{L}_a^0 \tau}  \tilde{\rho}_a \otimes \rho_{ss} d\tau \Big\} ~, \nonumber
\end{eqnarray}
where $\tilde{\mathcal{L}}_\textrm{int} \rho \equiv -i \lambda [\delta V (\hat{a}+\hat{a}^\dag),\rho]$, and $\delta V \equiv V-\langle V\rangle_{ss}$.  Due to limited time correlation, the upper bound of the integration is usually replaced by $t\rightarrow \infty$.  After applying Markov and rotating wave approximations, Eq.~(\ref{eq:ssmode_eq}) becomes
\begin{eqnarray}\label{eq:ssmode_eq2}
\dot{\tilde{\rho}}_a &=& \mathcal{L}^0_a \tilde{\rho}_a + \mathcal{L}^\mathcal{D}_a \tilde{\rho}_a -i \lambda^2 \textrm{Im}(S(\nu)+S(-\nu)) [\hat{a}^\dag \hat{a}, \tilde{\rho}_a]\nonumber \\ 
&&+2 \lambda^2 \Big(\textrm{Re}S(\nu) \mathcal{D}[\hat{a}]\tilde{\rho}_a + \textrm{Re}S(-\nu) \mathcal{D}[\hat{a}^\dag]\tilde{\rho}_a\Big)~.
\end{eqnarray}
The spectral function is given by
\begin{equation}\label{eq:ssmode_spectrum}
S(\nu) = \int^\infty_0 \textrm{Tr}\{\delta V U_q(t-t') \delta V \rho_{ss} \} e^{i \nu (t-t')} dt'~,
\end{equation}
where $U_q(t)$ is the evolution operator with respect to $\mathcal{L}_q$, i.e., $\partial_t U_q(t) = \mathcal{L}_q U_q(t)$.  The spectral function depends on only the qudit state, and can be calculated by quantum regression theorem \cite{book:GardinerZoller}.

By using Eq.~(\ref{eq:ssmode_eq2}), the mean phonon number of the mode, $\langle \tilde{n} \rangle = \textrm{Tr}\{\hat{a}^\dag \hat{a}\tilde{\rho}_a\}$, follows
\begin{equation}\label{eq:ssn_eq}
\langle \dot{\tilde{n}} \rangle = -\Gamma_c \langle \tilde{n} \rangle + \gamma \mathcal{N}~,
\end{equation}
where $\Gamma_c \equiv \lambda^2 \big(S(\nu)-S(-\nu)\big)+\gamma$ is the cooling rate, and $\gamma \mathcal{N} \equiv \lambda^2 S(-\nu)+\gamma N_\textrm{th}$ is the heating rate.  When the mode reaches a steady state, i.e., $\langle \dot{\tilde{n}} \rangle =0$, the final excitation of LD regime laser cooling is hence $n_\textrm{LD} \equiv \gamma \mathcal{N}/\Gamma_c$.  Here we have assumed that the steady state displacement $\alpha_\textrm{ss}$ is removed by coherent operation after cooling, otherwise the final excitation is modified by a high order factor as $n_f= n_\textrm{LD} + |\alpha_\textrm{ss}|^2$.

\section{Cooling in high temperature regime}

In the rest of this article, we retain the assumption of weak interaction strength, i.e., $\eta \ll 1$, but consider background temperatures that may exceed the LD regime, i.e., $\eta \sqrt{N_\textrm{th}} \gtrsim 1$.  In this situation, the interaction in Eq.~(\ref{eq:H_int}) cannot be treated perturbatively.  We observe that in laser cooling literatures, the mode state can always be viewed as an ensemble of coherent state with different displacement, i.e., 
\begin{equation}
\rho_a(t) =  \int P(\alpha,\alpha^\ast, t) |\alpha\rangle \langle \alpha| d^2\alpha~, 
\end{equation}
where the Glauber P function $P(\alpha,\alpha^\ast,t)$ is positive and smooth.
For this kind of states, the total state evolution can be obtained by studying the collective dynamics of each differently displaced sub-ensembles.


\begin{figure}
\begin{center}
\includegraphics{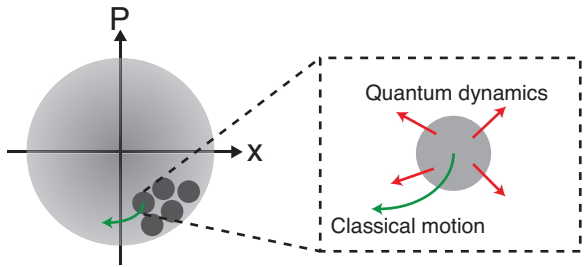}
\caption{ \label{fig:CQ} Illustration of our strategy to analyse laser cooling of a high temperature oscillator.  A high temperature P function is treated as an ensemble of sub-ensemble that have large classical displacements.  For each sub-ensemble, the effect of the mode-qudit interaction is treated perturbatively to obtain the cooling and the heating rate.}
\end{center}
\end{figure}

Our strategy based on this intuition is illustrated in Fig.~\ref{fig:CQ}.  A high temperature ensemble is equivalent to a collection of sub-ensembles with large displacement, i.e., 
\begin{equation}\label{eq:dss_total}
\rho(t) = \int p_c (\alpha, \alpha^\ast,t) D(\alpha)\tilde{\rho}_{\alpha}(t)D^\dag(\alpha) d^2\alpha~,
\end{equation}
where $p_c(\alpha, \alpha^\ast,t)$ is the classical probability of finding a sub-ensemble that is displaced by $\alpha$, and $\tilde{\rho}_\alpha$ is the mode-qudit state of that sub-ensemble.
The dynamics of each sub-ensemble obeys Eq.~(\ref{eq:master_LD}).  The mode-qudit interaction would generate a displacement-dependent force that in-turn affects the classical motion of the displacement.  
If the mode-qudit interaction is perturbative after subtracting the classical contribution, 
its effect on the mode state can be analysed by modifying the approach in the LD regime.  
Combining both effects for all sub-ensembles, we deduce a Fokker-Planck equation to describe the evolution of the Glauber P function of the total mode state.

A sufficient validity of our formalism is the possibility to express the total state as Eq.~(\ref{eq:dss_total}) and each sub-ensemble exhibits only perturbative mode-qudit interaction, i.e., $\eta \sqrt{\textrm{Tr}\{\hat{a}^\dag \hat{a}\tilde{\rho}_\alpha \}} \ll 1$.  This condition is self-consistently satisfied in our regime of interest where $\eta \ll 1$ and the Glauber P function is a probability of coherent states.

\subsection{Classical Displacement Damping \label{sec:classical_displacement}}

Here we just outline the key steps of our derivation; a detailed treatment is presented in Appendix \ref{app:derivation}.  
We first note that Eq.~(\ref{eq:dss_total}) is over-determined that we have the freedom to choose the implicit time dependence of $\alpha(t)$.  The choice determines the evolution of $p_c$ and $\tilde{\rho}_\alpha$; a proper choice of $\alpha(t)$ could simplify the problem.  Here we require $\alpha(t)$ to satisfy
\begin{equation}\label{eq:displacement_a}
\dot{\alpha}(t) + (i \nu +\gamma/2) \alpha (t) + i \lambda \langle V_\alpha (t)\rangle =0~,
\end{equation}
such that the relative displacement of the sub-ensemble vanishes, i.e., $\textrm{Tr}\{\hat{a}\tilde{\rho}_\alpha\}=0$.  

With this choice, $\alpha(t)$ can be recognized as the classical displacement of the sub-ensemble.  Eq.~(\ref{eq:displacement_a}) is hence the classical equation of motion of the sub-ensemble mode state.  Particularly, $i \lambda \langle V_\alpha(t) \rangle = i \lambda \textrm{Tr}\{V \tilde{\rho}_\alpha(t)\}$ accounts for the damping effect due to the mode-qudit interaction.  At the leading order of $\lambda$, the sub-ensemble qudit state, $\tilde{\rho}_{\alpha,q}\equiv \textrm{Tr}_a\{\tilde{\rho}_\alpha\}$, follows
\begin{equation}\label{eq:displacement_q}
\dot{\tilde{\rho}}_{\alpha,q} = \mathcal{L}_q \tilde{\rho}_{\alpha,q} - i \lambda \Big(\alpha(t)+\alpha^\ast (t) \Big) [V, \tilde{\rho}_{\alpha,q}]~.
\end{equation}

When comparing to the equilibrium condition in the LD regime, Eqs. (\ref{eq:ss_q}) and (\ref{eq:ss_a}), here the classical motion induces a time-dependent potential on the qudit.  Solving Eqs. (\ref{eq:displacement_a}) and (\ref{eq:displacement_q}) self-consistently is possible, but we further simplify the problem by considering that the displacement is dominated by the unperturbed harmonic motion, i.e., $\alpha(t)-\alpha_\textrm{ss} \approx r e^{-i \nu t}$, where $r$ is defined as the degree of motional excitation.  Then the motion-induced potential becomes periodic with the frequency $\nu$.

In analogy to the equilibrium assumption in the LD regime, we consider the qudit state converges to the dynamic steady state \cite{Ficek:2000tm}, which is defined as
\begin{equation}\label{eq:dss_q}
\rho_{ss} (t) = \sum_{n=-\infty}^\infty \rho_{ss}^{(n)} e^{i n \nu t}~,~\textrm{where  } \dot{\rho}_{ss}^{(n)} =0~.
\end{equation}
By inserting Eq.~(\ref{eq:dss_q}) into Eq.~(\ref{eq:displacement_q}), all coefficients $\rho_q^{(n)}$ can be obtained with the Floquet method \cite{book:Faisal}.

With the dynamic steady state, we can obtain the dynamic steady value of the qudit-induced damping effect in Eq.~(\ref{eq:displacement_a}), i.e., $i \lambda \langle V_\alpha(t) \rangle = i \lambda \sum_{n=-\infty}^\infty V_{n} e^{i n \nu t}$ and $\dot{V}_n=0$.  Particularly, the effect is dominated by the near-resonant term $i \lambda V_{-1} e^{-i \nu t}$.  As we will see, this is also the main contribution of the cooling rate for the total mode state.


\subsection{Quantum Dynamics Around Large Displacement \label{sec:quantum_diffusion}}

After subtracting the contribution of the classical motion, the mode-qudit interaction is perturbative, and so the quantum dynamics of the sub-ensemble can be analysed by modifying the techniques in the LD regime.  Our intuition is that if the mode state is initially localised within a small region in the phase space, for a sufficiently short time the coupling with the qudit and the environment will not project the state far from the region.  
In analogy to Eqs.~(\ref{eq:ssmode_eq2}) and (\ref{eq:ssmode_spectrum}), we find that the sub-ensemble mode state, $\tilde{\rho}_{\alpha,a}=\textrm{Tr}_q\{\tilde{\rho}_\alpha \} $, evolves as
\begin{eqnarray}\label{eq:dssmode_eq2}
%
%
&&\dot{\tilde{\rho}}_{\alpha,a} = \mathcal{L}^0_a \tilde{\rho}_{\alpha,a} + \mathcal{L}^\mathcal{D}_a \tilde{\rho}_{\alpha,a}  \\
&& -\lambda^2 \Big(S_\alpha(\nu, t) [\hat{a}+\hat{a}^\dag , \hat{a} \tilde{\rho}_{\alpha,a} ] +S_\alpha(-\nu, t) [\hat{a}+\hat{a}^\dag, \hat{a}^\dag \tilde{\rho}_{\alpha,a} ]\nonumber \\
&& - S_\alpha^\ast(-\nu, t) [\hat{a}+\hat{a}^\dag, \tilde{\rho}_{\alpha,a} \hat{a}] - S_\alpha^\ast(\nu, t) [\hat{a}+\hat{a}^\dag,  \tilde{\rho}_{\alpha,a} \hat{a}^\dag] \Big)~,\nonumber
\end{eqnarray}
where the time dependent spectral function $S_\alpha(\pm\nu, t)$ is
\begin{eqnarray}\label{eq:dssmode_spectrum}
&&S_\alpha(\pm\nu,t)  \\
&=& \int^t_0 \textrm{Tr} \{ \delta V_\alpha(t) U_q(t-t') \delta V_\alpha(t') \rho_{ss}(t') \} e^{\pm i \nu (t-t')} dt'~.\nonumber
\end{eqnarray}
We show in Appendix \ref{app:ssv} the techniques for obtaining the dynamic steady value of $S_\alpha$.

\subsection{Fokker-Planck Equation \label{sec:Fokker_Planck}}

The time evolution of the total mode state involves two parts: the redistribution of the classical displacement probability $p_c$ in Eq.~(\ref{eq:dss_total}), and the quantum evolution of the sub-ensemble mode states.  After combining both effects through the steps in Appendix \ref{app:FP}, the total mode state evolution can be described by a Fokker-Planck equation of the Glauber P function,
\begin{eqnarray} \label{eq:FP2}
\dot{P}&=& \frac{\partial}{\partial \alpha} \Big( (i \nu \alpha +\frac{\gamma}{2} \alpha +i \lambda \langle V_\alpha(t)\rangle)P\Big) - \lambda^2 \frac{\partial^2}{\partial \alpha^2} S_\alpha(-\nu, t) P \nonumber \\
&&+ \frac{1}{2} \frac{\partial^2}{\partial \alpha \partial \alpha^\ast} (\gamma N_\textrm{th}+2 \lambda^2 \textrm{Re} S_\alpha(-\nu,t))P +\textrm{h.c.}
\end{eqnarray}

Eq.~(\ref{eq:FP2}) can be simplified by making the rotating wave approximation with respect to $\nu$.
Furthermore, if the oscillator state is initially symmetric against the phase of $\alpha$, and the initial excitation is much larger than $|\alpha_\textrm{ss}|^2$, the P function would vary with respect to only $r$.  The Fokker-Planck equation for this component of P function is 
\begin{equation}\label{eq:FP}
\dot{P}(r,t)=\frac{1}{r}\frac{\partial}{\partial r} \left(\frac{r^2}{2}\Gamma_c(r) P(r) + \frac{r}{4}\frac{\partial}{\partial r} \gamma \mathcal{N}(r) P(r) \right)~,
\end{equation}
where the collective cooling rate is $\Gamma_c(r) \equiv \gamma+\textrm{Re}(2 i \lambda V_{-1}(r)/r - 2 \lambda^2 S_{-2}(-\nu,r)/r^2)$, and the collective heating rate is $\gamma \mathcal{N}(r) \equiv \gamma N_\textrm{th} + 2\lambda^2 \textrm{Re} (S_0(-\nu,r)-S_{-2}(-\nu,r))$;  the dynamic steady spectral functions are defined as $S_\alpha(-\nu,t)\equiv \sum_{n=-\infty}^\infty S_n(-\nu,r) e^{i n\nu t}$ and $\dot{S}_n=0$.

When $\dot{P}=0$, the steady state P function is given by
\begin{equation}\label{eq:Peq}
P_\textrm{eq}(r) = \frac{\mathcal{A}}{\gamma \mathcal{N}(r)} \exp\left( -\int_0^r \frac{2 r'\Gamma_c(r')}{\gamma \mathcal{N}(r')}dr' \right)~,
\end{equation}
where the normalization constant $\mathcal{A}$ guarantees $\int^\infty_0 r P_\textrm{eq}(r) dr=1$.  The final excitation is hence $n_f \equiv \int^\infty_0 r^3 P_\textrm{eq}(r) dr$.  We note that if the steady state displacement is neglected, the final excitation has to be modified as $n_f \rightarrow n_f + |\alpha_\textrm{ss}|^2$.

\section{Ladder system \label{sec:Ladder}}

\begin{figure}
\begin{center}
\includegraphics{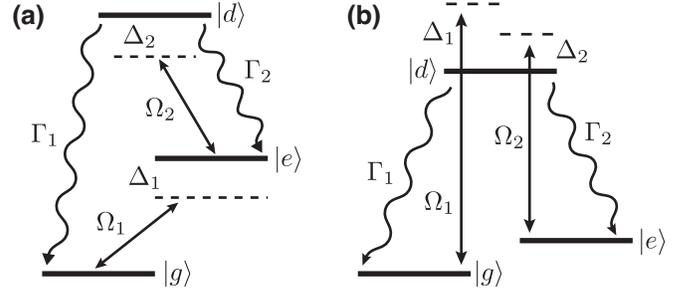}
\caption{Level diagram of (a) Ladder system, and (b) $\Lambda$ system. \label{fig:levels}}
\end{center}
\end{figure}

We apply the tools developed in the previous section to study Ladder-system cooling in the high temperature regime.  Specifically, we consider the following Hamiltonian and dissipative dynamics:
\begin{eqnarray}\label{eq:Ladder_H}
H_q &=& \Delta_1 \sigma_{ee} + (\Delta_1 + \Delta_2) \sigma_{dd} \nonumber \\
&&+\frac{\Omega_1}{2} (\sigma_{eg}+\sigma_{ge}) + \frac{\Omega_2}{2}(\sigma_{ed}+\sigma_{de})~, \\
V&=& \sigma_{ee}~,\\
\mathcal{L}^\mathcal{D}_q\rho &=& \Gamma_1 \mathcal{D}[\sigma_{gd}] \rho +\Gamma_2 \mathcal{D}[\sigma_{ed}] \rho~,
\end{eqnarray}
where $\sigma_{mn}\equiv |m\rangle \langle n|$ for qudit states $|m\rangle$ and $|n\rangle$.  The layout of the qudit levels is shown schematically in Fig.~\ref{fig:levels}(a).  For the purpose of cooling, the drive between $|g\rangle$ and $|e\rangle$ is red-detuned from the transition frequency, i.e., $\Delta_1>0$.  In each cycle of Ladder-system cooling, a phonon is absorbed during the transition from $|g\rangle$ to $|e\rangle$. Then the excitation at $|e\rangle$ is restored to $|g\rangle$ through an excitation to the fast-decaying state $|d\rangle$.

\begin{figure}
\begin{center}
\includegraphics{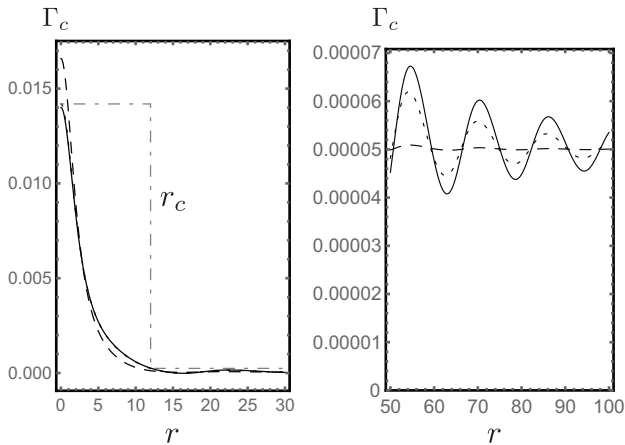}
\caption{ \label{fig:Ladder_G} Cooling rate of two Ladder systems with the same $\Delta_1=0.8\nu$, $\Delta_2=0$, $\Omega_1=0.6\nu$, $\Gamma_2=0$, $\eta=0.1$, $\gamma=5 \times 10^{-5}\nu$, but different in $\Gamma_1=2\nu$ and $\Omega_2=\sqrt{0.4}\nu$ (dashed), and $\Gamma_1=20\nu$ and $\Omega_2=2\nu$ (dotted).  The system parameters are chosen that both Ladder systems yield the same effective TLS.  Cooling rates of the corresponding TLS (solid) is also shown for comparison.  Left: small $r$ regime. $\Gamma_\textrm{toy}$ (dotdashed) is also shown.  Right: large $r$ regime.}
\end{center}
\end{figure}


The cooling rates produced by two typical Ladder-systems are shown in Fig.~\ref{fig:Ladder_G}.  In general, the cooling rate is high when the degree of motional excitation, $r$, is small.  As $r$ increases to the order of $r\gtrsim 1/\eta$, the cooling rate is reduced by orders of magnitude, and eventually dominated by the bare damping rate $\gamma$.  

\begin{figure}
\begin{center}
\includegraphics{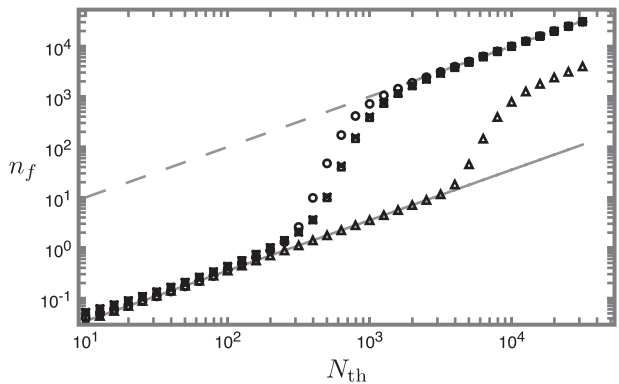}
\caption{ \label{fig:Ladder_nf} Final excitation of Ladder system cooling with the same system parameters as in Fig.~\ref{fig:Ladder_G}.  Except in the transient region during transition, the results of $\Gamma_1=2\nu$ (circle) and $\Gamma_1=20\nu$ (cross) match that of the effective TLS (square).  Solid line shows the final excitation predicted by LD regime analysis, and dashed line shows the equilibrium excitation without cooling.  The final excitation of the toy model (triangle) is also shown for reference.}
\end{center}
\end{figure}

The final excitation of Ladder-system cooling is shown in Fig.~\ref{fig:Ladder_nf}.  
For a wide range of $N_\textrm{th}$, which could exceed the LD criterion in Eq.~(\ref{eq:LD_criterion}), the final excitation matches the prediction of LD regime analysis, $n_\textrm{LD}$.  Above some critical background temperature, cooling becomes inefficient and the final excitation rises until reaching $N_\textrm{th}$.  As a side note, the critical transition is not determined by the initial excitation of the mode, because the steady state given by Eq.~(\ref{eq:Peq}) is unique.


The criticality of the final excitation is also observed in TLS cooling \cite{Rabl:2010cm}.  We attribute the criticality to the reduction of cooling rate at large $r$.  
To illustrate our idea, we study a toy model that the cooling rate is a step function, i.e., $\Gamma_\textrm{toy}(r<r_c)=\Gamma_c(0)$ and $\Gamma_\textrm{toy}(r>r_c)=\Gamma_c(r_c)$ for some  `jump' point $r_c$.  $\Gamma_\textrm{toy}$ is by construction the upper-bound of $\Gamma_c$ of a Ladder-system.  If a final excitation transition appears in the toy model, so does it happens in the realistic case.  For simplicity, we assume the heating rate is constant as $\gamma N_\textrm{th}$, which is a good approximation for a high temperature background or a mode with a low Q factor.

The analytical expression of the toy model final excitation is
\begin{equation}\label{eq:T_toy}
n_f = \frac{n_\textrm{LD}^2 + e^{-r_c^2/n_\textrm{LD}} \big(n_+ (n_++r_c^2) - n_\textrm{LD} (n_\textrm{LD} + r_c^2) \big) }{n_\textrm{LD} + e^{-r_c^2/n_\textrm{LD}} n_+}~,
\end{equation}
where $n_\textrm{LD}\equiv\gamma N_\textrm{th}/\Gamma_c(0)$; $n_+\equiv\gamma N_\textrm{th}/\Gamma_c(r_c) \gg n_\textrm{LD}$ for a sufficiently large $r_c$.  As shown in Fig.~\ref{fig:Ladder_nf}, the final excitation of the toy model exhibits a transition.  Specifically, if $n_\textrm{LD} \ll r_c^2$, the second parts in both the denominator and the numerator are suppressed by an exponentially small factor $e^{-r_c^2/n_\textrm{LD}}$.  Therefore, the final excitation can be approximated by $n_f \approx n_\textrm{LD}$, which matches the prediction of the LD regime analysis.  On the other hand, if $n_\textrm{LD} \gtrsim r_c^2$, the second parts become dominant and the final excitation is $n_f \approx n_+$.  In this regime cooling becomes inefficient.  

The critical value of $N_\textrm{th}$ in Ladder-system or TLS cooling is closely related to the system parameters.  Nevertheless, we can estimate the value of $r_c$ by considering at which $r$ the LD model becomes invalid.  We consider the interaction picture with respect to the motion-induced potential; the state in this picture is transformed as $\rho_{\alpha,q}' \equiv \exp(i 2 \eta r \sin \nu t \sigma_{ee} ) \tilde{\rho}_{\alpha,q} \exp(-i 2 \eta r \sin \nu t \sigma_{ee})$.  After applying the Jacobi-Anger expansion \cite{book:Cantrell}, Eq.~(\ref{eq:displacement_q}) becomes
\begin{equation}
\dot{\rho}_{\alpha,q}' = -i[H'_q,\rho_{\alpha,q}'] + \mathcal{L}^\mathcal{D}_q\rho_{\alpha,q}'~,
\end{equation}
where 
\begin{eqnarray}
H'_q &=& \Delta_1 \sigma_{ee} + (\Delta_1 + \Delta_2) \sigma_{dd}  \\
&&+\frac{\Omega_1}{2} \sum_{n=-\infty}^\infty J_n(2\eta r)(e^{i n \nu t}\sigma_{eg}+e^{-i n \nu t}\sigma_{ge}) \nonumber \\
&&+ \frac{\Omega_2}{2}\sum_{n=-\infty}^\infty J_n(2\eta r) (e^{i n \nu t}\sigma_{ed}+e^{-i n \nu t}\sigma_{de})~;\nonumber
\end{eqnarray}
and $J_n$ are Bessel functions of the first kind.  The transformation changes only the unperturbed Hamiltonian of the qudit, but not the qudit-induced damping effect in Eq.~(\ref{eq:displacement_a}), i.e., $ \textrm{Tr}\{V\rho'_{\alpha,q} \}=\langle V_\alpha (t) \rangle$.

When comparing with Eq.~(\ref{eq:Ladder_H}), the drives in the original picture are effectively suppressed by a factor of $J_0(2 \eta r)$.  At the same time, additional sideband drives are induced with a magnitude proportional to $J_n(2 \eta r)$.  We can claim that the cooling mechanism in the LD regime is no longer a proper description of the system dynamics when $r_c \sim 1/2 \eta$, at where the magnitude of the original drives is significantly reduced and the sideband drives become non-negligible.  Combining with the intuition obtained from the toy model, the final excitation matches the prediction of the LD regime analysis only if, 
\begin{equation}\label{eq:mLD_criterion}
\eta \sqrt{n_\textrm{LD}} \ll 1~,
\end{equation}
which is a weaker condition than Eq.~(\ref{eq:LD_criterion}).  Eq.~(\ref{eq:mLD_criterion}) can also be understood as a self-consistent condition: at the steady state the mode-qudit interaction has to be a perturbative factor so that the LD regime analysis is valid.

Finally, we discuss the quality of the effective TLS engineering.  In the LD regime, an effective TLS can be engineered by appropriately driving a Ladder-system and adiabatically eliminating the fast-decaying state $|d\rangle$ \cite{Marzoli:1994tz}.  This matches our observation when $r$ is small: as shown in Fig.~\ref{fig:Ladder_G}, the cooling rates of both Ladder-system are close to that of the TLS with the same effective system parameters.  On the other hand, the Ladder-system cooling rate is much smaller in magnitude than the TLS cooling rate when $r$ is large.  We attribute this effect to the shift of $|e\rangle$ state level by the classical motion-induced potential, so that the $|e\rangle \leftrightarrow |d\rangle$ transition becomes off-resonant.  This suppresses the restoration of the qudit state to $|g\rangle$, and hence reduces the cooling rate.
We also note that, at any $r$ a Ladder-system with a larger $\Gamma_1$ acts as a better approximation to the effective TLS.  
This is because the off-resonance of $|e\rangle \leftrightarrow |d\rangle$ transition is less important if the line-width of $|d\rangle$ is large.

The cooling rate discrepancy can only be experienced by a sub-ensemble if it is at large $r$.  Therefore, the final excitation can be affected only if the total mode state consists of a significant probability of highly excited sub-ensembles, in other words, only if cooling is inefficient.  The idea can be observed in Fig.~\ref{fig:Ladder_nf} that, the final excitation of Ladder-system and TLS cooling mismatch only when $n_f$ transits from $n_\textrm{LD}$ to $N_\textrm{th}$.  
Nevertheless, our main interest is the cases that cooling is efficient.  Then the final excitation is mainly determined by the cooling rate at the small $r$ regime.  In this regime, the cooling rate, and hence the final excitation, produced by the Ladder-systems and TLS are close.  Our conclusion is that even though the background temperature is beyond the LD regime, if TLS cooling is found to be efficient with certain system parameters, the effective parameters can be engineered by driving a Ladder-system without compromising the cooling performance.

\section{$\Lambda$ system \label{sec:Lambda}}

The Hamiltonian and the dissipative dynamics of a $\Lambda$-system are
\begin{eqnarray}\label{eq:Lambda_H}
H_q &=& -\Delta_1 \sigma_{gg} -\Delta_2 \sigma_{ee} \nonumber \\
&&+\frac{\Omega_1}{2} (\sigma_{gd}+\sigma_{dg}) + \frac{\Omega_2}{2}(\sigma_{ed}+\sigma_{de})~,\\
V&=& -\sigma_{gg}+\sigma_{ee}~,\\
\mathcal{L}^\mathcal{D}_q\rho &=& \Gamma_1 \mathcal{D}[\sigma_{gd}] \rho +\Gamma_2 \mathcal{D}[\sigma_{ed}] \rho~.
\end{eqnarray}
The layout of the qudit levels is shown schematically in Fig.~\ref{fig:levels}(b).  $\Lambda$-system has been extensively studied not only because it appears in various quantum systems, but it also exhibits EIT that an excitation from $|g\rangle$ is dependent of the drive between other states.  EIT has a variety of applications ranging from storing light to improving photon-photon interaction \cite{Fleischhauer:2005da}.  Particularly it can be applied in laser cooling \cite{Morigi:2000wv,Roos:2000ch,Morigi:2003cv,Lin:2013be, Lechner:2016EIT}.  

The operational condition of EIT cooling is remarkably different from TLS cooling: the metastable states are driven in blue-detuning, instead of red-detuning, from the resonant transition frequency, i.e., $\Delta<0$.  Besides, efficient cooling requires quantum coherence between $|g\rangle$ and $|e\rangle$.  
EIT cooling has several advantages over TLS cooling schemes.  For examples, a lower final excitation can be attained due to the darkness of the steady state \cite{Morigi:2000wv,Morigi:2003cv}, and multiple oscillator modes can be cooled simultaneously \cite{Lin:2013be}.



\begin{figure}
\begin{center}
\includegraphics{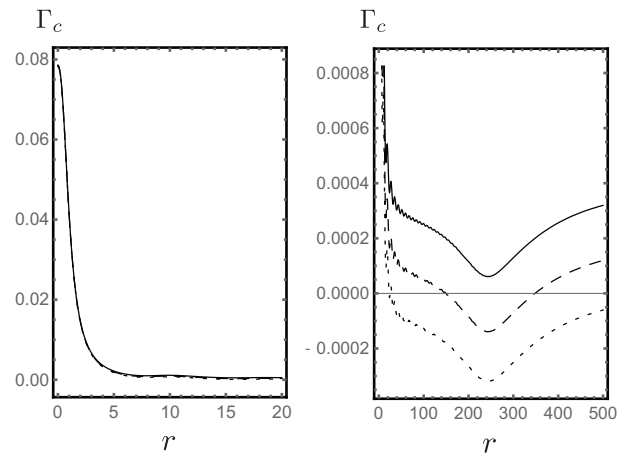}
\caption{ \label{fig:Lambda_G} EIT Cooling rate a $\Lambda$ system with $\Delta_1=\Delta_2=-50\nu$, $\Omega_1=\Omega_2=\nu$, $\Gamma_1=\Gamma_2=10\nu$, $\lambda=0.1\nu$.  The only difference is the Q factor of the oscillator, which are $\gamma=2\times 10^{-5}\nu$ (dotted), $\gamma=2\times 10^{-4}\nu$ (dashed), $\gamma=4\times 10^{-4}\nu$ (solid). Left: Small $r$ regime.  Right: Large $r$ regime.}
\end{center}
\end{figure}

The EIT cooling rate with a typical $\Lambda$-system is plotted in Fig.~\ref{fig:Lambda_G} for different $r$.  Similar to the Ladder-system and TLS cooling in the small $r$ regime, the EIT cooling rate decreases as $r$ increases.  The crucial difference appears in the large $r$ regime.  Instead of reducing to negligible in magnitude, the qudit contribution of the cooling rate, $\Gamma_c-\gamma$, remains negative for a wide range of $r$.  This is because the motion-induced potential shifts the energy levels, so the Raman transition between $|g\rangle$ and $|e\rangle$ is no longer in resonance.  Then both $|g\rangle$ and $|e\rangle$ are driven in blue-detuned frequency, which is a process that would increase the motional excitation.

\begin{figure}
\begin{center}
\includegraphics{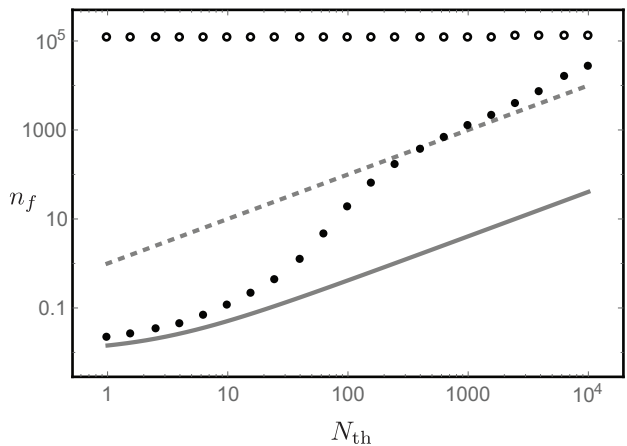}
\caption{ \label{fig:Lambda_nf} Final excitation of two EIT-cooled system with the same system parameters as in Fig.~\ref{fig:Lambda_G}.  $\gamma=4\times 10^{-4}\nu$ (solid circle) corresponds to the case of usual cooling behaviour, and $\gamma=2 \times 10^{-4}\nu$ (hollow circle) is dominated by the lasing effect.  $n_\textrm{LD}$ (solid line) and $N_\textrm{th}$ (dashed line) are also shown for comparison.}
\end{center}
\end{figure}

The regime of negative cooling rate causes dramatic effect on the steady state.  In Fig.~\ref{fig:Lambda_nf}, we show two typical behaviours of final excitation produced by EIT cooling.  Below some threshold Q factor, EIT cooling behaves similarly as Ladder-system and TLS cooling: the final excitation is close to $n_\textrm{LD}$ for small $N_\textrm{th}$, and cooling becomes inefficient when $N_\textrm{th}$ is large.  On the other hand, when the Q factor is above some threshold, the final excitation is much higher than, and only slightly depending on, the background temperature.
Our results contradict to the common belief that a higher Q factor would lead to a lower final excitation.

\begin{figure}
\begin{center}
\includegraphics{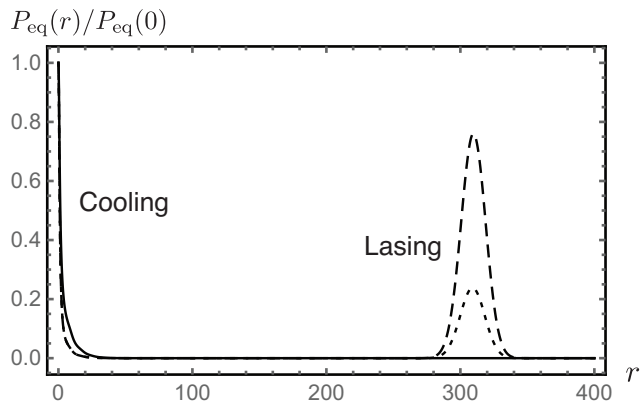}
\caption{ \label{fig:Lambda_P} Equilibrium radial P function of EIT cooling at background temperature $N_\textrm{th}=300$.  System parameters are the same as in Fig.~\ref{fig:Lambda_G} with $\gamma=4\times 10^{-4}\nu$ (solid), $\gamma=2.56\times 10^{-4}\nu$ (dashed), $\gamma=2.55\times 10^{-4}\nu$ (dotted).  For better comparison, the P functions are normalised as unity at $r=0$.}
\end{center}
\end{figure}

The criticality of Q factor can be understood from the two types of steady state P function that can be produced by EIT cooling.  As shown in Fig.~\ref{fig:Lambda_P}, for a sufficiently small Q factor the negativity of $\Gamma_c-\gamma$ is compensated, so $\Gamma_c$ is positive for every $r$.  Therefore, the mode is always cooled and the steady state P function is a single peak around $r=0$.  

On the other hand, a large Q factor results in negative $\Gamma_c$ at some $r$.  The motion of the sub-ensemble will be amplified if at its displacement $\Gamma_c$ is negative.  
For this range of Q factor, in addition to the usual cooling peak at $r=0$, the steady state P function consists of a large $r$ peak centred at where the cooling rate vanishes, i.e., $\Gamma_c \approx 0$.  For a larger Q factor, the large $r$ peak contributes more significantly and eventually dominates the behaviour of the steady state.  In analogy to the equilibrium photon state of a laser \cite{book:ScullyZubairy}, here the large $r$ peak is a signature of the phonon lasing effect due to the blue sideband drives from $|g\rangle$ and $|e\rangle$.

In conventional laser cooling, the cooling drives are applied to the qudit until the mode has reached the steady state.  Following this convention, EIT cooling can reduce the motional excitation of a mode only if its Q factor is sufficiently small that the lasing effect is suppressed.  
Nevertheless, EIT cooling is still applicable to cool a mode with large Q factor, if the operation is halted at a transient stage before the steady state is reached.  

\begin{figure}
\begin{center}
\includegraphics{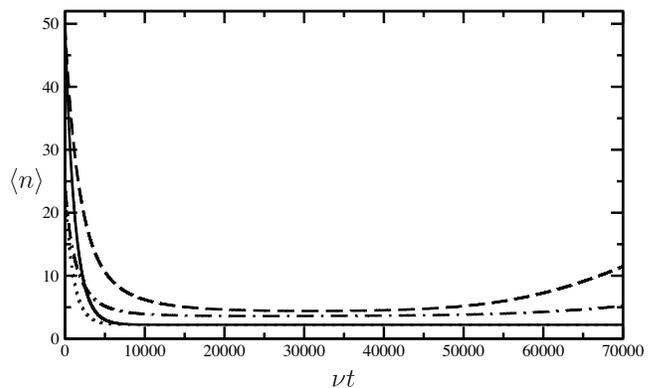}
\caption{ \label{fig:Lambda_t} Time variation of mean motional excitation $\langle n\rangle$.  The system parameters are the same as Fig.~\ref{fig:Lambda_G} except $\gamma$ and $N_\textrm{th}$.  The background heating rate is fixed as $\gamma N_\textrm{th}=2\times 10^{-2}\nu$.  For $\gamma=4\times 10^{-4}$, where cooling dominates, the mean excitation converges to the final temperature $n_f\approx 2.2$ for both initial temperature $\langle n(0)\rangle =50$ (solid) and $\langle n(0)\rangle =25$ (dotted).  For $\gamma=1\times 10^{-4}$, where lasing dominates, the mean excitation exhibits transient minimum at $\langle n\rangle \approx 4.4$ for $\langle n(0)\rangle =50$ (dashed), and $\langle n \rangle \approx 3.6$ for $\langle n(0)\rangle =25$ (dot-dashed).}
\end{center}
\end{figure}

To illustrate the idea, we simulate the mode state time evolution by numerically integrating Eq.~(\ref{eq:FP}).  As shown in Fig.~\ref{fig:Lambda_t}, the mean excitation is reduced significantly after EIT cooling is applied, and stays at the minimum for a fairly long time before it rises.  The intuition here is that $\Gamma_c$ is positive around $r=0$, at where cooling is efficient.  If the initial excitation of the mode is sufficiently low, with a high probability a sub-ensemble is found within the efficient cooling regime.  The emergence of phonon lasing effect requires the total mode state to diffuse, due to the heating rate $\gamma \mathcal{N}$, from the regime of positive to negative $\Gamma_c$.  If the diffusion process requires a longer time than cooling, at some transient stage the mode is efficiently cooled.

We note that the minimum transient excitation achieved by this method depends on the initial excitation.  In contrast, the final temperature of conventional cooling is independent of the initial excitation, due to the uniqueness of the steady state.  Our results suggest that pre-cooling is important for EIT cooling to be efficient.



\section{Conclusion \label{sec:conclusion}}

In this article, we study the laser cooling of a mechanical oscillator with a dissipative three level system.  
Under a background temperature exceeding the Lamb-Dicke regime, we propose a formalism to separate the large classical motion from the quantum dynamics of the mode, and calculate the total cooling and heating rate in a self-consistent way.  

In Ladder-system cooling, the cooling rate generally reduces by orders of magnitude when the displacement reaches $|\alpha|\sim 1/\eta$.  We show that the reduction of cooling rate causes a critical transition of the steady state excitation, which indicates that cooling is no longer efficient when the background temperature is beyond some critical value.  We provide a simple criterion to estimate the regime of efficient cooling.  In this regime, a Ladder-system can be used to engineer the system parameters of an effective two-level system without compromising the cooling performance.


When EIT cooling is operated with a $\Lambda$-system, its high temperature performance is remarkably different from the Ladder-system and the TLS cooling.  For a wide range of displacement, the EIT cooling rate could be negative, so that the steady state simultaneously exhibits both cooling and lasing effects.  
We suggest that successful EIT cooling would require either a small Q factor to suppress the lasing effect, or terminating the operation at a transient stage.  In the latter case, the minimum transient temperature is dependent of the initial excitation of the mode.

We end this article with a discussion about possible strategies for better cooling performance.  In TLS-like cooling scheme, the remaining question is to minimise the steady state excitation at the LD regime.  This could be done by exploring the optimal system parameters and the internal level layout, e.g. Ladder-system or V-system.
On the other hand, optimising EIT cooling would be more complicated because the minimum excitation state may be transient.  Possible strategies includes applying TLS-like cooling to pre-cool the mode before applying EIT cooling, or dynamically varying the system parameters so that the majority of sub-ensembles experience positive cooling rate throughout the process.  
Overall, our work provides intuitions and tools for designing and analysing laser cooling schemes for high temperature oscillators.

H.-K. L. would like to acknowledge the support from the Croucher Foundation.  This work was supported by an Alexander von Humboldt Professorship.

\appendix

\section{Derivation of Fokker Planck equation \label{app:derivation}}

This appendix provides details of the tedious steps that enter the derivation of the Fokker-Planck equation, Eq.~(\ref{eq:FP}), for the radial P function.

\subsection{Master equation in harmonic reference frame}

We first derive Eq.~(\ref{eq:dssmode_eq2}) that describes the quantum dynamics of a sub-ensemble with classical displacement $\alpha(t)$.  
If the displacement satisfies Eq.~(\ref{eq:displacement_a}), then the sub-ensemble mode-qudit state obeys
\begin{equation}\label{eq:master_DLD}
\dot{\tilde{\rho}}_\alpha = \mathcal{L}_q\tilde{\rho}_\alpha+\mathcal{L}_1(t) \tilde{\rho}_\alpha +  \tilde{\mathcal{L}}_\textrm{int}(t)\tilde{\rho}_\alpha + \mathcal{L}^0_a \tilde{\rho}_\alpha+ \mathcal{L}^\mathcal{D}_a \tilde{\rho}_\alpha~,
\end{equation}
which is the same as Eq.~(\ref{eq:master_LD}) except the addition of a time dependent potential induced by the classical displacement, i.e., $\mathcal{L}_1(t) \rho \equiv - i \lambda \big(\alpha(t)+\alpha^\ast (t) \big) [V, \rho]$, and the mode-qudit interaction is replaced by $\tilde{\mathcal{L}}_\textrm{int}(t)\rho \equiv -i \lambda [\delta V_\alpha(t) (\hat{a}+\hat{a}^\dag),\rho]$, where $\delta V_\alpha(t) \equiv V-\langle V_\alpha (t)\rangle$.  If the unperturbed evolution of the qudit is much faster than that of the mode and the mode-qudit interaction, then at the leading order the qudit is disentangled from the mode state and its dynamics follows Eq.~(\ref{eq:displacement_q}).

We then assume the classical motion is approximately harmonic, $\alpha(t)-\alpha_\textrm{ss} \approx r e^{-i \nu t}$.  The assumption is valid in our case because the bare damping rate and the mode-qudit interaction strength are much smaller than $\nu$.  Here we have neglected the phase of the displacement for simplicity; it can be trivially added back by redefining the time reference of the oscillating phase.  

Under the harmonic assumption, the motion-induced potential becomes periodic with $\nu$.  We then conduct Floquet analysis to obtain the dynamic steady state of the qudit in Eq.~(\ref{eq:dss_q}) \cite{book:Faisal}.  Following the same idea as the LD regime analysis, our aim is to adiabatically eliminate the contribution of the qudit and to construct an equation for the mode state only.  

Specifically, we define the projection operators
\begin{equation}
\mathcal{P}X(t) \equiv \textrm{Tr}_q\{X(t)\} \otimes \rho_\textrm{ss} (t)~,
\end{equation}
and $\mathcal{Q}=\mathbb{I}-\mathcal{P}$ for any density operator $X$ in the mode-qudit Hilbert space.  After applying the projection onto Eq.~(\ref{eq:master_DLD}), we have
\begin{eqnarray}
\textrm{Tr}_q\{ \dot{\tilde{\rho}} \} \otimes \rho_\textrm{ss}(t) = (\mathcal{L}^0_a + \mathcal{L}^\mathcal{D}_a ) \mathcal{P}\tilde{\rho} + \mathcal{P}\mathcal{L}_\textrm{int}(t) \mathcal{Q} \tilde{\rho}&& \\
\dot{\mathcal{Q} \tilde{\rho}} = (\mathcal{L}_q + \mathcal{L}_1(t) + \mathcal{L}^0_a + \mathcal{L}^\mathcal{D}_a ) \mathcal{Q}\tilde{\rho} +\mathcal{Q} \mathcal{L}_\textrm{int}(t) \mathcal{P} \tilde{\rho}&&,
\end{eqnarray}
where we have used the identities $[\mathcal{L}^0_a, \mathcal{P}] = [\mathcal{L}^\mathcal{D}_a, \mathcal{P}]=0$, $\mathcal{P}\tilde{\mathcal{L}}_\textrm{int}(t)\mathcal{P}\rho=0$, $\mathcal{P}\mathcal{P}=\mathcal{P}$, $\textrm{Tr}_q\{\mathcal{Q}\rho\}=0$, and $\textrm{Tr}_q\{\mathcal{L}_q\rho\}=0$.

In analogy to Eq.~(\ref{eq:ssmode_eq}), each sub-ensemble mode state follows
\begin{eqnarray}\label{eq:dssmode_eq}
\dot{\tilde{\rho}}_{\alpha,a}(t) &=& \mathcal{L}^0_a \tilde{\rho}_{\alpha,a}(t) + \mathcal{L}^\mathcal{D}_a \tilde{\rho}_{\alpha,a}(t)  \\
&&+ \textrm{Tr}_q \Big\{ \int_0^t \tilde{\mathcal{L}}_\textrm{int}(t) U_q(t-t')U_a(t-t') \nonumber \\
&& \times \tilde{\mathcal{L}}_\textrm{int}(t') \big(\tilde{\rho}_{\alpha,a}(t') \otimes \rho_{ss}(t')\big) d t' \Big\} ~,~\nonumber
\end{eqnarray}
where the evolution operators are defined as $\rho_q(t) \equiv U_q(t-t') \rho_q(t')$ for $\rho_q$ satisfying Eq.~(\ref{eq:displacement_q}); $U_a(t-t') \tilde{\rho}_{\alpha,a}(t') \approx e^{-i \nu t \hat{a}^\dag \hat{a}} \tilde{\rho}_{\alpha,a}(t') e^{i \nu t \hat{a}^\dag \hat{a}}$ as $\mathcal{L}^\mathcal{D}_a$ is assumed to be at the second order of $\lambda$.  

Before moving forward, we note several important differences between Eqs.~(\ref{eq:ssmode_eq}) and (\ref{eq:dssmode_eq}).  First, the time independent steady quantities in the LD regime are substituted by the dynamic steady quantities that are varying as time.  Therefore the time ordering in the integral becomes important.  Second, the upper bound of the integration cannot be replaced by $t\rightarrow\infty$.  As we will see, the $t$ dependence in the integration bound is necessary for obtaining the correct evolution of the spectral function.

In analogy to the derivation of Eq.~(\ref{eq:ssmode_eq2}), we consider the evolution of the mode is dominated by the free evolution, i.e., $U_a(t-t')\tilde{\rho}_{\alpha,a}(t') \approx \tilde{\rho}_{\alpha,a}(t)$, then Eq.~(\ref{eq:dssmode_eq}) can be written as Eq.~(\ref{eq:dssmode_eq2}).  


\subsection{Steady state values \label{app:ssv}}

Solving Eq.~(\ref{eq:dssmode_eq2}) requires the dynamic steady values of the qudit state $\rho_\textrm{ss}(t)$ and the spectral function $S_\alpha(\pm\nu,t)$.  Here we discuss the techniques for computing these values.

The dynamics of the qudit state following Eq.~(\ref{eq:displacement_q}) can be solved by considering the evolution of the generalized bloch vector, $\langle\bm{\sigma}(t) \rangle \equiv \textrm{Tr}\{\bm{\sigma}\tilde{\rho}_{\alpha,q}(t) \}$, where $\{\bm{\sigma}_i\}$ is a complete set of operator in the qudit state space.  In our three-level system analysis, we pick $\bm{\sigma} \equiv (\sigma_{gg}~ \sigma_{ee}~\sigma_{ge}~\sigma_{eg}~\sigma_{dd}~\sigma_{gd}~\sigma_{dg}~\sigma_{ed}~\sigma_{de})^\textrm{T}$.  The Bloch vector version of Eq.~(\ref{eq:displacement_q}) is given by
\begin{equation}\label{eq:matrix_eq1}
\dot{\langle \bm{\sigma}\rangle} = \Big(\mathbf{M} -i \lambda \big(\alpha(t)+\alpha^\ast (t)\big) \mathbf{V} \Big) \cdot \langle\bm{\sigma}\rangle~,
\end{equation}
where the time independent matrix elements are defined as $\textrm{Tr}\{\bm{\sigma}_i \mathcal{L}_q\tilde{\rho}_{\alpha,q} \} \equiv \sum_j \mathbf{M}_{i j} \langle \bm{\sigma}_j\rangle$, and $\textrm{Tr} \{ [\bm{\sigma}_i , V ] \tilde{\rho}_{\alpha,q}\}\equiv \sum_j\bm{V}_{ij}\langle \bm{\sigma}_j\rangle$.

After making the harmonic approximation on the displacement, finding the dynamic steady state is equivalent to finding the solution of Eq.~(\ref{eq:matrix_eq1}) that satisfies
\begin{equation}\label{eq:dss_state}
\langle\bm{\sigma}(t)\rangle = \sum_{n=-\infty}^{\infty} \vec{\sigma}^{(n)} e^{i n \nu t}~,
\end{equation}
where $\dot{\vec{\sigma}}^{(n)}=0$.

Eq.~(\ref{eq:matrix_eq1}) is not easy to solve because $\mathbf{M}$ is singular.  The singularity originates from the conservation of the trace, $\textrm{Tr}\{\tilde{\rho}_{\alpha,q}(t)\}=1$.  Such condition can be removed from the equation by transforming the Bloch vector as $\mathcal{T}\cdot \langle\bm{\sigma}\rangle = (1~\langle\tilde{\bm{\sigma}}\rangle^\textrm{T})^\textrm{T}$, and then considering only the second to ninth entries.  We specifically consider $\mathcal{T}$ is a 9x9 matrix that corresponds to the transformation $\tilde{\bm{\sigma}}_1 = \sigma_{gg}+\sigma_{ee}+\sigma_{dd}$, $\tilde{\bm{\sigma}}_2 = -\sigma_{gg}+\sigma_{ee}$, $\tilde{\bm{\sigma}}_5 =\sigma_{ee}-\sigma_{dd}$, and $\tilde{\bm{\sigma}}_i=\bm{\sigma}_i$ otherwise.  After the transformation, we have
\begin{equation}\label{eq:matrix_eq2}
\langle \dot{\tilde{\bm{\sigma}} }\rangle = \Big(\tilde{\mathbf{M}} -i \lambda r\big(e^{-i\nu t}+e^{i\nu t}\big) \tilde{\mathbf{V}} \Big) \cdot \langle\tilde{\bm{\sigma}}\rangle + \bm{u}~,
\end{equation}
where $\tilde{\mathbf{M}}=\mathcal{R}\mathcal{T}(\mathbf{M}-i2\lambda \textrm{Re}(\alpha_\textrm{ss})\mathbf{V})\mathcal{T}^{-1}\mathcal{R}^\textrm{T}$, $\tilde{\mathbf{V}}=\mathcal{R}\mathcal{T} \mathbf{V} \mathcal{T}^{-1}\mathcal{R}^\textrm{T}$, and $\bm{u}=\mathcal{R}\mathcal{T} \mathbf{M}\mathcal{T}^{-1}\cdot (1~\bm{0}_8^\textrm{T})^\textrm{T}$.  The truncation matrix is defined as $\mathcal{R}\equiv (\bm{0}_8~\mathbb{I}_8)$, where $\bm{0}_8$ is a 8x1 zero vector; $\mathbb{I}_n$ is a $n$x$n$ identity matrix.  

Substituting the dynamic steady Bloch vector in Eq.~(\ref{eq:dss_state}) into Eq.~(\ref{eq:matrix_eq2}) and matching the components with the same frequency, we can see the solution obeys
\begin{equation}
i n \nu \vec{\sigma}^{(n)} = \tilde{\mathbf{M}}\cdot \vec{\sigma}^{(n)} -i \lambda r \tilde{\bm{V}} \cdot (\vec{\sigma}^{(n+1)}+\vec{\sigma}^{(n-1)}) + \delta_{n,0} \bm{u}~,
\end{equation}
where $\delta_{n,0}$ is the Kronecker delta.  By assuming $\vec{\sigma}^{(n)}$ vanishes for a large enough $n$, the dynamic steady $\langle \tilde{\bm{\sigma}} \rangle$, and hence $\langle \bm{\sigma}\rangle$, can be obtained efficiently by the continued fraction method \cite{Rabl:2010cm}.

The spectral function can be solved in a similar way.  We first note that the spectral function in Eq.~(\ref{eq:dssmode_spectrum}) is identical to 
\begin{equation}
S_\alpha(\pm\nu,t) = \int^t_0 \textrm{Tr} \{V U_q(t-t') \delta V_\alpha(t') \rho_{ss}(t') \} e^{\pm i \nu (t-t')} dt'~.
\end{equation}
We then define the spectral vector as
\begin{equation}\label{eq:spectral_eq}
\bm{S}(\pm \nu,t) = \int^t_0 \textrm{Tr} \{\bm{\sigma} U_q(t-t') \delta V_\alpha(t') \rho_{ss}(t') \} e^{\pm i \nu (t-t')} dt'~.
\end{equation}
If the spectral vector is obtained, the spectral function can be obtained by picking the specific components, i.e., $S_\alpha=\bm{S}_2$ for Ladder-system, and $S_\alpha=-\bm{S}_1+\bm{S}_2$ for $\Lambda$-system.

Differentiating Eq.~(\ref{eq:spectral_eq}) with respect to $t$ and applying the quantum regression theorem \cite{book:GardinerZoller}, we have
\begin{eqnarray}\label{eq:spectral_eq2}
\dot{\bm{S}}(\pm\nu, t) &=& \Big(\pm i \nu \mathbb{I}_9 + \mathbf{M} -i \lambda \big(\alpha(t)+\alpha^\ast (t)\big) \mathbf{V}\Big) \cdot \bm{S}(\pm\nu, t) \nonumber \\
&&+ \textrm{Tr}\{\bm{\sigma}\delta V_\alpha(t) \rho_\textrm{ss}(t) \}~.
\end{eqnarray}
We note that the last term, which is essential to recover the LD regime result at the limit $\lambda\rightarrow 0$, would be missing if the integral upper bound in Eq.~(\ref{eq:dssmode_spectrum}) is taken as $t\rightarrow \infty$.

Under the harmonic approximation of the displacement, the dynamic steady spectral function can be obtained from the solution of Eq.~(\ref{eq:spectral_eq2}) that satisfies
\begin{equation}\label{eq:dss_spectral}
\bm{S}(\pm \nu, t) =\sum_{n=-\infty}^\infty \vec{S}_n(\pm \nu) e^{i n \nu t},
\end{equation}
and $\dot{\vec{S}}_n(\pm \nu)=0$.  Nevertheless, solving Eq.~(\ref{eq:spectral_eq2}) also suffers from the problem of the singularity of $\mathbf{M}$.  We again extract the singular component by the transformation $\mathcal{T}\cdot\mathbf{S} = (0~\tilde{\mathbf{S}}^\textrm{T})^\textrm{T}$.
In the transformed basis, Eq.~(\ref{eq:spectral_eq3}) can be written as
\begin{eqnarray}\label{eq:spectral_eq3}
\dot{\tilde{\bm{S}}}(\pm\nu, t) &=& \Big(\pm i \nu \mathbb{I}_8 + \tilde{\mathbf{M}} -i \lambda r \big(e^{i \nu t}+ e^{-i \nu t}\big) \tilde{\mathbf{V}}\Big)  \nonumber \\
&&\cdot \tilde{\bm{S}}(\pm\nu, t)+\textrm{Tr}\{\tilde{\bm{\sigma}}\delta V_\alpha(t) \rho_\textrm{ss}(t) \}~.
\end{eqnarray}
 Substituting Eq.~(\ref{eq:dss_spectral}) into Eq.~(\ref{eq:spectral_eq3}), the dynamic steady solution obeys
 \begin{eqnarray}
 i n \nu \vec{S}_n(\pm \nu) &=& (\pm i \nu  \mathbb{I}_8 + \tilde{\mathbf{M}})\cdot \vec{S}_n(\pm \nu) \\
 &&-i \lambda r\tilde{\mathbf{V}}\cdot \big(\vec{S}_{n+1}(\pm \nu)+\vec{S}_{n-1}(\pm \nu)\big) + \vec{v}_n~, \nonumber
 \end{eqnarray}
where $\textrm{Tr}\{\tilde{\bm{\sigma}}\delta V(t) \rho_\textrm{ss}(t) \}\equiv\sum_{n=-\infty}^\infty \vec{v}_n e^{i n \nu t}$.  By assuming $\vec{S}_n$ vanishes for a sufficiently large $n$, the solution can be obtained by the continued fraction method \cite{Rabl:2010cm}.

Before moving forward, we note that the above method can reproduce the LD regime steady state and spectral function by setting $\lambda \rightarrow 0$ and considering only the non-oscillating component, i.e., $n=0$.  

\subsection{Fokker-Planck equation \label{app:FP}}

Differentiating Eq.~(\ref{eq:dss_total}) and tracing out the qudit contribution, we get 
\begin{eqnarray}\label{eq:drhodt_total}
\dot{\rho}_a(t) &=& \int \dot{p}_c(\alpha,\alpha^\ast,t) D(\alpha)\tilde{\rho}_{\alpha,a}(t) D^\dag(\alpha) d^2\alpha \\
&&+ \int p_c(\alpha,\alpha^\ast,t) D(\alpha)\dot{\tilde{\rho}}_{\alpha,a}(t) D^\dag(\alpha) d^2 \alpha ~. \nonumber
\end{eqnarray}
The dynamics of the total P function consists of two parts: the redistribution of the displacement due to the classical motion of each sub-ensemble, and the quantum dynamics of the mode state of each sub-ensemble.  Our aim is to obtain an equation that relates those time derivatives to the operations involving only $\alpha$ and $\alpha^\ast$.

For the first part in Eq.~(\ref{eq:drhodt_total}), $p_c$ evolves according to the continuity equation of probability density \cite{book:LandauLifshitz}:
\begin{equation}
\dot{p}_c = - \dot{\alpha} \frac{\partial p_c}{\partial \alpha} - \dot{\alpha}^\ast\frac{\partial p_c}{\partial \alpha^\ast} -\Big(\frac{\partial \dot{\alpha}}{\partial \alpha} + \frac{\partial \dot{\alpha}^\ast}{\partial \alpha^\ast} \Big) p_c~,
\end{equation}
where $\dot{\alpha}=-(i\nu+\gamma/2)\alpha - i \lambda\langle V_\alpha(t)\rangle$ according to Eq.~(\ref{eq:displacement_a}).  

For the second part, the time derivatives of the sub-ensemble P function, $P_\alpha$, can be expressed as
\begin{equation}
\dot{\tilde{\rho}}_{\alpha,a}(t) =  \int \dot{P}_\alpha (\beta,\beta^\ast, t) |\beta \rangle \langle \beta | d^2\beta~.
\end{equation}
By applying standard conversion rules \cite{book:GardinerZoller} to Eq.~(\ref{eq:dssmode_eq2}), we can obtain an dynamic equation for each sub-ensemble P function.

Combining both effects, the total mode state varies as
\begin{widetext}
\begin{eqnarray}\label{eq:combine_eq}
\dot{\rho}_a &=& \int 
\Big( \frac{\partial}{\partial \alpha} \big(i \nu \alpha + \frac{\gamma}{2}+i \lambda \langle V_\alpha(t) \rangle\big) p_c(\alpha,\alpha^\ast,t) +\frac{\partial}{\partial \alpha^\ast} \big(-i \nu \alpha^\ast + \frac{\gamma}{2}-i \lambda \langle V_\alpha(t) \rangle\big) p_c(\alpha,\alpha^\ast,t) \Big) \nonumber \\
&&\times P_\alpha(\beta,\beta^\ast,t) |\alpha+\beta\rangle\langle\alpha+\beta| d^2\alpha d^2\beta \nonumber \\
&&+\int  \Big( -i (\nu + \lambda^2 \textrm{Im}\big(S_\alpha(\nu,t)+S_\alpha(-\nu,t)\big))\Big(-\frac{\partial}{\partial \beta}\beta P_\alpha(\beta,\beta^\ast,t) +\frac{\partial}{\partial \beta^\ast}\beta^\ast P_\alpha(\beta,\beta^\ast,t)  \Big)\nonumber \\
&&+(\frac{\gamma}{2}+\lambda^2 \textrm{Re}\big(S_\alpha(\nu,t)-S_\alpha(-\nu,t)\big))\Big(\frac{\partial}{\partial \beta}\beta P_\alpha(\beta,\beta^\ast,t) +\frac{\partial}{\partial \beta^\ast}\beta^\ast P_\alpha(\beta,\beta^\ast,t)  \Big) \nonumber \\
&&+ \lambda^2 \big(-S_\alpha(\nu,t)+S_\alpha^\ast (-\nu,t) \big) \frac{\partial}{\partial \beta^\ast} \beta P_\alpha (\beta,\beta^\ast,t) + \lambda^2 \big(S_\alpha(-\nu,t)-S_\alpha^\ast (\nu,t) \big) \frac{\partial}{\partial \beta} \beta^\ast P_\alpha (\beta,\beta^\ast,t) \nonumber \\
&& + \big(\gamma N_\textrm{th}+2\lambda^2 \textrm{Re}S_\alpha(-\nu,t)\big) \frac{\partial^2}{\partial \beta \partial \beta^\ast} P_\alpha (\beta,\beta^\ast,t) -\lambda^2 S_\alpha(-\nu,t)\frac{\partial^2}{\partial \beta^2}P_\alpha(\beta,\beta^\ast,t) -\lambda^2 S_\alpha^\ast(-\nu,t)\frac{\partial^2}{\partial \beta^{\ast 2}}P_\alpha(\beta,\beta^\ast,t) \Big) \nonumber \\
&&\times p_c(\alpha,\alpha^\ast,t) |\alpha+\beta\rangle\langle\alpha+\beta| d^2\alpha d^2\beta~.
\end{eqnarray}
\end{widetext}

Eq.~(\ref{eq:combine_eq}) is redundant as it involves the operations of both $\alpha$ and $\beta$.  The redundancy arises because the sub-ensemble states are not specified.  In other words, there could be different choices of sub-ensemble P function that constitutes the same total P function:
\begin{equation}
P(\alpha,\alpha^\ast,t) = \int p_c(\alpha,\alpha^\ast,t) P_\alpha(\beta-\alpha,\beta^\ast - \alpha^\ast,t) d^2 \beta~.
\end{equation}

In fact, this is a degree of freedom that we could employ to simplify Eq.~(\ref{eq:combine_eq}).  We recall the assumption that at any time $t$ the cooling oscillator state can be treated as an ensemble of coherent state, i.e., every sub-ensemble P function is momentarily a Dirac delta function, $P_\alpha(\beta,\beta^\ast,t)=\delta(\beta)\delta(\beta^\ast)$.  Then at that moment $t$ the P function coincides with the classical probability,
\begin{eqnarray}
\rho_a(t)&=&\int p_c(\alpha,\alpha^\ast,t)\delta(\beta)\delta(\beta^\ast) |\alpha+\beta \rangle\langle \alpha+\beta |d^2\alpha d^2\beta  \nonumber\\
&=& \int p_c(\alpha,\alpha^\ast,t)|\alpha\rangle \langle \alpha| d^2\alpha \nonumber \\
&=&\int P(\alpha,\alpha^\ast,t)|\alpha\rangle \langle \alpha| d^2\alpha~. 
\end{eqnarray}
After tedious but straight forward steps, Eq.~(\ref{eq:combine_eq}) can be written as the form
\begin{equation}
\dot{\rho}_a(t) = \int \dot{P}(\alpha,\alpha^\ast,t) |\alpha\rangle \langle \alpha | d^2 \alpha~,
\end{equation}
where $\dot{P}$ is given by Eq.~(\ref{eq:FP2}).





\bibliographystyle{phaip}
\pagestyle{plain}
\bibliography{cooling_bib}

\end{document}